\documentclass[aps, prc,
reprint,
longbibliography, % Show paper titles in citations
nofootinbib,
preprintnumbers,
superscriptaddress
]{revtex4-2}

\usepackage[flushleft]{threeparttable} % http://ctan.org/pkg/threeparttable
\usepackage{booktabs,caption}
\usepackage[inline]{enumitem}

\usepackage[colorlinks, urlcolor=blue, citecolor=blue, menucolor=.,
  anchorcolor=., linkcolor=., runcolor=.]{hyperref}
\makeatletter
\def\@bibdataout@aps{%
\immediate\write\@bibdataout{%
@CONTROL{%
apsrev41Control%
\longbibliography@sw{%
    ,author="08",editor="1",pages="1",title="0",year="1"%
    }{%
    ,author="08",editor="1",pages="1",title="",year="1"%
    }%
  }%
}%
\if@filesw \immediate \write \@auxout {\string \citation {apsrev41Control}}\fi
}
\makeatother

\usepackage{setspace}

\usepackage{amsfonts}
\usepackage{amsmath}
\usepackage{amssymb}
\usepackage{physics}
\usepackage{subcaption}

\DeclareFontFamily{OT1}{pzc}{}
\DeclareFontShape{OT1}{pzc}{m}{it}{<-> s * [1.10] pzcmi7t}{}
\DeclareMathAlphabet{\mathpzc}{OT1}{pzc}{m}{it}

\usepackage{booktabs}
\AtBeginDocument{
\heavyrulewidth=.08em
\lightrulewidth=.05em
\cmidrulewidth=.03em
\belowrulesep=.65ex
\belowbottomsep=0pt
\aboverulesep=.4ex
\abovetopsep=0pt
\cmidrulesep=\doublerulesep
\cmidrulekern=.5em
\defaultaddspace=.5em
}

\usepackage{comment} % comment environment
\usepackage[inline]{enumitem}
\usepackage[nowatermark]{fixmetodonotes}
\usepackage{graphicx}
\usepackage{isotope}
\usepackage{siunitx}

\DeclareSIUnit\ton{t}
\DeclareSIUnit\parsec{pc}
\DeclareSIUnit[number-unit-product = ]\percent{\char`\%}

\usepackage{orcidlink}

\begin{document}

\preprint{FERMILAB-PUB-26-0117-CSAID-T}

\title{ %First Quasi-Elastic Neutrino Scattering Parameter Estimation Using Simulation-Based Inference
First Estimation of Model Parameters for Neutrino-Induced Nucleon Knockout
Using Simulation-Based Inference
% Inference-Driven Tuning of Neutrino Interaction Model Parameters
% Simulation-based inference for neutrino interaction model parameter tuning
}
\author{Karla Tame-Narvaez\,\orcidlink{0000-0002-2249-9450}\,}
\affiliation{Fermi National Accelerator Laboratory, Batavia, Illinois 60510 USA}
\author{Steven Gardiner\,\orcidlink{0000-0002-8368-5898}\,}
\affiliation{Fermi National Accelerator Laboratory, Batavia, Illinois 60510 USA}
\author{Aleksandra \'Ciprijanovi\'c\,\orcidlink{0000-0003-1281-7192}\,}
\affiliation{Fermi National Accelerator Laboratory, Batavia, Illinois 60510 USA}
\affiliation{Department of Astronomy and Astrophysics, University of Chicago, Chicago, IL 60637}
\affiliation{NSF-Simons AI Institute for the Sky (SkAI), 172 E. Chestnut St., Chicago, IL 60611}
\author{Giuseppe Cerati\,\orcidlink{0000-0003-3548-0262}\,}
\affiliation{Fermi National Accelerator Laboratory, Batavia, Illinois 60510 USA}

\date{\today}

\begin{abstract}
To enable an accurate determination of oscillation parameters, accelerator-based
neutrino experiments require detailed simulations of nuclear interaction
physics in the GeV regime. While substantial effort from both theory and
experiment is currently being invested to improve the fidelity of these
simulations, their present deficiencies typically oblige experimental
collaborations to resort to empirical tuning of simulation model parameters.
As the precision requirements of the field continue to become more stringent,
machine learning techniques may provide a powerful means of handling
corresponding growth in the complexity of future neutrino interaction model
tuning exercises. To study the suitability of simulation-based inference (SBI)
for this physics application, in this paper we revisit a tuned configuration
of the GENIE neutrino event generator that was originally developed by the
MicroBooNE collaboration. Despite closely reproducing the adopted values of
four physics parameters when confronted with the tuned cross-section
predictions as input, we find that our trained SBI algorithm prefers modestly
different values (within MicroBooNE's assigned uncertainties) and achieves
slightly better goodness-of-fit when inference is run on the experimental data
set originally used by MicroBooNE. We also find that our
trained algorithm can create a fair approximation of an alternative neutrino
scattering simulation, NuWro, that shares only a subset of its physics model
parameters with GENIE.
\end{abstract}


\maketitle

%%%%%%%%%%%%%%%%%%%%%%%%%%%%%%%%%%%%%%%%%
\section{Introduction}
%%%%%%%%%%%%%%%%%%%%%%%%%%%%%%%%%%%%%%%%%

Obtaining a precise and comprehensive understanding of neutrino oscillations is
a goal at the forefront of worldwide investments in high-energy physics.
Next-generation neutrino experiments based at accelerator facilities, including
Hyper-Kamiokande~\cite{Hyper-Kamiokande:2018ofw} and the Deep Underground
Neutrino Experiment (DUNE)~\cite{DUNE1, DUNE2}, will require percent-level
control of systematic uncertainties in order to deliver their flagship
oscillation results~\cite{NuSTEC:2017hzk}. Among the sources of uncertainty for
these efforts, some of the most challenging are those arising from imperfect
modeling of neutrino-nucleus interactions at the relevant GeV energy scale. Due
to the broad range of energies generated by present neutrino beam facilities,
as well as the complexity of inferring the incident neutrino energy from each
detected event, oscillation measurements can be sensitive to various details of
the interaction model used in the analysis. While considerable research effort
is being directed towards improving the interaction simulations used by the
neutrino community, modeling deficiencies remain problematically large even in
state-of-the-art codes. As a result, experimental neutrino collaborations
typically rely on data-driven mitigation strategies in order to achieve the
needed measurement precision.

A common approach for addressing the mismodeling of neutrino scattering has been
empirical tuning of simulation parameters to cross-section data. Using
predictions from the popular GENIE event generator~\cite{genie2010, genie2021}
as a starting point, the MINERvA~\cite{MINERvA:2019kfr},
NOvA~\cite{NOvA:2020rbg}, and MicroBooNE~\cite{microboonegenietune}
collaborations have all developed their own tuned neutrino interaction models
in recent years. This experiment-specific work has been pursued in parallel
with more generic tuning studies carried out by the GENIE collaboration
itself~\cite{GENIE:2021zuu, GENIE:2021wox, GENIE:2022qrc, GENIE:2024ufm}.
Although such tuning must be executed carefully, both to ensure robust
oscillation results~\cite{Coyle:2025xjk} and to avoid obscuring potential
signatures of exotic physics~\cite{Coyle:2022bwa}, this general technique has
provided a successful foundation for numerous modern results in neutrino
physics. In light of its established track record and the notable difficulty of
comprehensive, \textit{a priori} modeling of nuclear effects at GeV energies,
empirical tuning of neutrino interaction simulations is likely to continue as
standard practice in the field. However, as the precision goals of future
oscillation analyses become increasingly stringent, the complexity of related
tuning campaigns will necessarily grow in response. More detailed input data
and a larger number of tuned parameters may both be required in the future to
achieve satisfactory simulation performance, and these constraints will in turn
lead to greater computational challenges.

To address these anticipated future needs, artificial intelligence and machine
learning (AI/ML) techniques may provide especially powerful solutions. As an
example particularly well-suited to this use case, we consider simulation-based
inference (SBI), which has gained popularity across scientific disciplines over
the last several years. SBI includes a broad group of methods that leverage
simulators as a part of the statistical inference procedure, where the goal is
to infer the likelihood or posterior distribution of underlying parameters for
a given experiment or observation~\cite{CB2020, DB2025}. SBI methods are
valuable in situations where forward simulations are available, but the
likelihood is unknown or intractable (e.g., when using a complex model with
hundreds or even thousands of parameters), and hence explicit likelihood
calculations are not feasible.

Traditional SBI methods have been present for some time, and include for
example, Approximate Bayesian Computation~\cite[ABC;][]{Rubin1984} and
Approximate Frequentist Computation~\cite[AFC;][]{brehmer2018guide},  which are
closely related to the traditional template histogram and kernel density
estimation approaches. More recently, SBI algorithms suitable for
high-dimensional parameter spaces have been developed, which utilize deep neural networks as surrogates for modeling the conditional probability densities of the likelihood or posterior distribution~\cite{CB2020}.

Deep learning-based SBI models have already been successfully used in different
areas of physics. In astrophysics and cosmology, these methods have been used
for the dark matter substructure inference in galaxy-galaxy strong
lenses~\cite{Brehmer_2019, Coogan2022, Anau2023}, inference of strong lensing
parameters~\cite{Legin2021, WagnerCarena2022, Poh2025}, inference of galaxy
properties from spectra~\cite{Khullar2022}, cosmology inference from galaxy
cluster abundance~\cite{Reza2022}, inference of the Hubble constant from binary
neutron star mergers~\cite{GF2021}, etc. In collider physics~\cite{BR2021}, SBI
has been used for constraining the Higgs potential for di-Higgs
production~\cite{MN2024}, searching for CP violation in leptonic WH
production~\cite{BM2024}, measuring QCD splittings~\cite{BB2021}, etc.

Building on these precedents from other areas of physics, as well as recent
work on SBI for modeling the hardware response of the JUNO neutrino
detector~\cite{Gavrikov:2025rps}, in this article we examine the suitability of
SBI as a tool for neutrino interaction model tuning. For this application, SBI
provides an efficient alternative to conventional methods, as it offers
amortized inference, i.e., after the upfront cost of training the model is
paid, inference can be performed in seconds. To provide a realistic context for
our exploration of SBI-based tuning, we revisit the tuned GENIE model adopted
by the MicroBooNE collaboration~\cite{microboonegenietune}. Using a software
framework called NUISANCE~\cite{nuisance}, MicroBooNE collaborators adjusted
four GENIE simulation parameters in comparison to an external neutrino
cross-section measurement reported by the T2K experiment~\cite{T2K:2016jor}.
Our choice of this \textit{MicroBooNE Tune} is motivated by its relative
simplicity among other representative examples in the field, its continuing use
in the latest MicroBooNE analyses\footnote{While the MicroBooNE Tune
interaction model remains unaltered, additional data-driven, analysis-specific
model constraints are applied in some cases.}~\cite{MicroBooNE:2025nll,
MicroBooNE:2025aiw}, and recent scrutiny from the community into the robustness
of its predictions and underlying tuning methodology~\cite{Wolfs:2025ofb}.

In the remainder of this article, we demonstrate that our SBI algorithm can
\begin{enumerate*}[label=(\arabic*)] \item infer the correct parameter values
when confronted with mock data generated using the MicroBooNE Tune
configuration of GENIE,\footnote{See also our earlier demonstration
of this capability in Ref.~\cite{Tame-Narvaez:2025pwg}.}
\item describe the T2K data with equal or better
precision than the MicroBooNE Tune while overcoming issues with Peelle's Pertinent Puzzle~\cite{ppp, fruhwirth2012peelle} encountered in the original fitting procedure, and \item adjust the
input GENIE model to approximate the predictions of NuWro~\cite{golan2012},
an alternative neutrino event generator that seeks to model
the same physics processes as GENIE. \end{enumerate*}
These results pioneer the use of SBI to obtain event generator parameter values
from actual experimental data sets, thus establishing a novel approach,
potentially more powerful and scalable, for tuning neutrino interaction
simulations. In Section~\ref{data} we describe our training data, and in
Section~\ref{methods} we document the SBI algorithm, its neural network
architecture, and our training procedure. In Section~\ref{performance}, we
benchmark the performance of our SBI algorithm, and in Section~\ref{results} we
show the results of our fits to T2K data and a NuWro prediction. Finally, in
Section~\ref{summary}, we provide a summary and conclusions from our work.

%%%%%%%%%%%%%%%%%%%%%%%%%%%%%%%%%%%%%%%%%
\section{Training Data}
\label{data}
%%%%%%%%%%%%%%%%%%%%%%%%%%%%%%%%%%%%%%%%%

We simulate neutrino–nucleus interactions using the GENIE framework~\cite{genie2010,genie2021}.
Within this interaction model, we vary the four parameters that were adjusted in the MicroBooNE Tune, namely:
\begin{itemize}
    \item $\theta_1$ (\texttt{MaCCQE}). This parameter is used in the calculation of the charged-current quasielastic (CCQE) cross section. It represents the mass that appears in a dipole parameterization of the axial-vector form factor of the nucleon. It has an important impact on the normalization CCQE events, as well as the dependence of the cross section on the negative square of the four-momentum transfer $Q^2$.

    \item$\theta_2$
    (\texttt{NormCCMEC}). This parameter scales the overall normalization of the charged-current multi-nucleon interaction cross-section arising from meson exchange currents (CCMEC).

    \item $\theta_3$ (\texttt{XSecShape\_CCMEC}). This parameter controls the shape of the CCMEC cross section within the two-dimensional phase space of leptonic energy transfer and three-momentum transfer. As the parameter value moves between zero and one, GENIE interpolates between the shape predictions from two different cross-section models. Parameter values outside of this range are invalid.

    \item $\theta_4$ (\texttt{RPA\_CCQE}). This parameter controls the strength of the Random Phase Approximation (RPA) correction applied to CCQE processes. It has an important impact on both the normalization and the shape of the cross section in $Q^2$.
\end{itemize}

Further information on the implementation and physical meaning of these parameters is given in the original MicroBooNE Tune publication~\cite{microboonegenietune}. To provide sufficient coverage of the parameter space in the vicinity of the best-fit values from the MicroBooNE Tune, we created an ensemble of configurations in which the four parameters were independently sampled from uniform distributions with the following ranges:\footnote{Values of these parameters are consistently reported in this article in physical units. Different units are used internally by the GENIE implementation.}
\[
\begin{aligned}
\theta_1 &\in [0.528, 1.39]~\text{GeV}, \quad
\theta_2 \in [0,\, 3.0], \\
\theta_3 &\in [0.0, 1.0], \quad
\theta_4 \in [0.0, 1.5].
\end{aligned}
\]

%\footnote{The transformations used to go from GENIE to physical units are:
%$\theta_1 = M_A(1+0.03 * \theta_1^{\text{GENIE}})$, where $M_A$ is the axial mass, which is a parameter in the dipole model of the nucleon axial-vector form factor,  $\theta_2= 1 + 0.5* \theta_2^{\text{GENIE}}$,  $\theta_3=\theta_3^{\text{GENIE}}$ and   $\theta_4 = 1 - \theta_4^{\text{GENIE}}$}:
%\[
%\begin{aligned}
%\theta_1^{\text{GENIE}} &\in [-15, 15], \quad
%\theta_2^{\text{GENIE}} \in [-2,\, 4], \\
%\theta_3^{\text{GENIE}} &\in [0.0, 1.0], \quad
%\theta_4^{\text{GENIE}} \in [-0.5, 1].
%\end{aligned}
%\]

For each configuration of the parameters, we processed the output of GENIE with
NUISANCE~\cite{nuisance} to produce a 58-bin histogram representing a
theoretical prediction that can be directly compared to the ``Analysis~I'' T2K
neutrino interaction data set reported in Ref.~\cite{T2K:2016jor}. For the
original MicroBooNE Tune, this procedure was used together with a subsequent
likelihood fit to the measured histogram to obtain best-fit parameter values.
In this study, we created a training set of $171,747$ configurations and a test
set of 1,000 independent configurations to test our SBI workflow. Data used in this project is publicly available on Zenodo.\footnote{\url{https://zenodo.org/uploads/18747090}}

%%%%%%%%%%%%%%%%%%%%%%%%%%%%%%%%%%%%%%%%%
\section{Methods}
\label{methods}
%%%%%%%%%%%%%%%%%%%%%%%%%%%%%%%%%%%%%%%%%

In traditional explicit Bayesian inference, the posterior $p(\theta | x)$ is given by Bayes' theorem as:
\begin{equation}
\begin{gathered}
p(\theta | x) = \frac{p(x | \theta) * p(\theta)}{p(x)},
\end{gathered}
\end{equation}
where $x$ denotes the data vector, $\theta$ is the set of model parameters to be inferred,
$p(x | \theta)$ is the likelihood, $p(\theta)$ is the prior on the parameters, and $p(x)$ is the evidence (or marginal likelihood). However, in case of many complex scientific problems, the likelihood function is unknown or computationally intractable (due to high model complexity), making explicit likelihood calculation impossible. To address these challenges, SBI (also referred to in some papers as Implicit Likelihood Inference (ILI) or Likelihood-Free Inference (LFI)) methods have gained popularity in recent years. The most recent and advanced SBI methods utilize deep learning models to directly create surrogates for the likelihood~\cite{Papamakarios2018SequentialNL}, likelihood ratio~\cite{Hermans2019} or posterior distribution~\cite{Papamakarios2016,Greenberg2019} in a given inference problem. These models are trained using pairs of input model parameters $\theta$ and model outputs, i.e., simulated data $x$. Due to high model capacity and amortized inference, deep-learning-based SBI methods opened doors for inference with uncertainty quantification even in high-dimensional parameter spaces~\cite{CB2020}.

%%%%%%%%%%%%%%%%%%%%%%%%%%%%%%%%%%%%%%%%%
 \subsection{Neural Posterior Estimation}
%%%%%%%%%%%%%%%%%%%%%%%%%%%%%%%%%%%%%%%%%

In this work, we focus on direct inference of parameter posteriors from the data $p(\theta | x)$, i.e., Neural Posterior Estimation (NPE). We perform one round of the NPE training based on the NPE-A method from~\cite{Papamakarios2016}. Note that there are two extensions of this method (NPE-B~\cite{NPE-B} and NPE-C~\cite{Greenberg2019}), which utilize sequential training. These Sequential NPE (SNPE) approaches utilize multiple rounds of training, where the posteriors inferred in one round are used as proposal priors for the next round. Unfortunately, in Ref.~\cite{Hermans2021ATC} it was shown that sequential approaches often produce overconfident posteriors. To utilize the full benefits of amortized inference and to avoid potential issues with our inferred posteriors, we limit our initial analysis to only one round of NPE training.

In NPE, a neural network with weights $\phi$, is trained on $M$ parameter-data pairs $\{\theta_j, x_j\}^M_{j=1}$ from the simulator, to approximate the posterior $p(\theta|x).$ Note that in our case each $\theta_j$ is actually a set of four parameters $\theta_j=\{\theta_j^i\}^4_{i=1}$ described in Section~\ref{data}, but for notation simplicity we omit this additional index here. If we denote this neural network-based approximate posterior with $q_{\phi}(\theta|x)$, we can then train the neural network by maximizing the average log-likelihood with respect to $\phi$:

\begin{equation}
\label{eq:gen_likelihood}
    \mathcal{L}(\phi) = \frac{1}{M}\sum^M_i\log q_{\phi}(\theta_i|x_i).
\end{equation}
If $q_{\phi}(\theta|x)$ is a sufficiently flexible model, and $M \rightarrow \infty$, this approximate posterior should converge to the true posterior $p(\theta|x)$ when the average log-likelihood is maximized.  In other words, by using the negative average log-likelihood as a loss function, we can train the neural network on the training parameter-data pairs to approximate the true posterior.

To build our NPE, we utilize Masked Autoregressive Flows (MAFs)~\cite{papamakarios2018maskedautoregressiveflowdensity}, due
their state-of-the-art flexibility in modeling complex density distributions. MAFs represent a fusion between two families of deep learning density estimators-- autoregressive models~\cite{Uria2016NeuralAD} and normalizing
flows~\cite{JimenezRezende2015}. Autoregressive density estimators utilize the fact that any $n$-dimensional probability density $p(\vb*{v})$, where $\vb*{v} \equiv [v_1,...,v_n]$, can be factorized into a product of 1-dimensional
conditional probability distributions by using the chain rule. On the other hand, normalizing flows focus on transforming samples from a simple base density, like a standard normal distribution, into samples from a more complex target density by using invertible transformations. Hence, a particular type of autoregressive models, which are parameterized as Gaussian conditional probabilities, can also be thought of as belonging to the normalizing flow family of models. In this case, one autoregressive block will contain conditional probabilities which can be written as:
\begin{equation}
\label{eq:autogauss}
p(v_k | \vb*{v}_{1:k-1}) = {\cal N} (v_k | \mu_k, (\mathrm{exp}\, \sigma_k)^2),
\end{equation}
where $p(v_k | \vb*{v}_{1:k-1})$ is the $k$-th conditional, conditioned on the subset of the vector containing data points with indices less than $k$, i.e., $\vb*{v}_{1:k-1}$. Additionally, $\mu_k$ and $\sigma_k$ are the mean and log standard deviation of the Gaussian that $v_k$ is sampled from, and are themselves scalar functions of previous data points $\vb*{v}_{1:k-1}$:

\begin{align}
    \label{eq:autoscalar1}
     & \mu_k = f_{\mu_k}(\vb*{v}_{1:k-1}), \\
     \label{eq:autoscalar2}
     & \sigma_k = f_{\sigma_k}(\vb*{v}_{1:k-1}).
\end{align}

To allow for more flexibility, we can then create stacks of these autoregressive model blocks, such that samples of the output density of the last block are used as the input for the next block. In this case, the initial input density (the standard normal distribution) is progressively transformed into samples from increasingly more complex distributions, which are then used as input densities for the following autoregressive model blocks to produce even more complex output densities, until we reach the final target data distribution $\vb*{v}$. This stacking greatly increases the flexibility of the density model, allowing application to complex scientific problems.

Finally, MAFs use a Masked Autoencoder for Distribution Estimation~\cite[MADE;][]{GG2015} to encode the autoregressive model blocks described previously. The main advantage of MADE over other autoregressive model architectures is that it avoids recursive computation. Each MADE block represents a neural network which utilizes binary masks applied to some of the network weights, such that in each of the autoregressive blocks, the functions $f_{\mu_{k}}$ and $f_{\sigma_{k}}$ for all $k$-s can be calculated efficiently in a single forward pass. For more technical details about MADE, we refer the reader to Ref.~\cite{GG2015}, and about MAFs to Ref.~\cite{papamakarios2018maskedautoregressiveflowdensity}.

%%%%%%%%%%%%%%%%%%%%%%
\subsection{Model Architecture and Training}
%%%%%%%%%%%%%%%%%%%%%%

The aim of this work is to infer the underlying set of parameters used
within GENIE from a histogram generated according to the approach
described in Section~\ref{data}. We utilize the SBI framework developed by
\texttt{mackelab}\footnote{\url{https://github.com/sbi-dev/sbi}}, implemented in
\texttt{python}. Our SBI model takes as input the four physics parameters discussed in Section~\ref{data},
$\theta_i$, along with their associated histograms, $x$, to learn the inverse
mapping from histograms to parameters, i.e., we aim to directly learn the posteriors of the $\theta_i$ parameters $p(\theta_i | x)$.

To improve efficiency in learning the posterior distribution, we employ a three-layer fully connected embedding network with ReLU activations after the first two layers. The hidden layers contain $45$ and $32$ neurons, respectively, progressively reducing the input dimensionality from $58$ to a $20$-dimensional output embedding \footnote{We note that when the full 58-dimensional histograms are used as input, the model tends to become overconfident; however, reducing the input dimensionality to fewer than 16 parameters leads to a loss of precision.}. These embeddings are then used as inputs to the NPE, for which we use the MAF architecture with three MADE blocks and $55$ hidden features in each. Both the embedding network and the MAF are trained together to allow the embedding network to learn the most informative summaries, which will lead to the best posterior predictions. The code used in this project can be found in our GitHub repository.\footnote{\url{https://github.com/karlaTame/SBI_parameter_tuning}}

During SBI model training, we use mini-batches of size $512$, a learning rate of $10^{-2}$, and a training/validation split of $90\%/10\%$ to monitor and mitigate overfitting. We also use an early stopping criterion, with patience of $50$ epochs. The training converges in an average of 300 epochs, running approximately 7 minutes in a regular CPU environment. We use a fixed random seed set to $555$ for reproducibility, but confirm that changing the seed does not present significant changes to the SBI performance, calibration, nor the final fits.

%%%%%%%%%%%%%%%%%%%%%%%%%%%%%%%%%%%%%%%%%
\section{Model performance and calibration}
\label{performance}
 %%%%%%%%%%%%%%%%%%%%%%%%%%%%%%%%%%%%%%%%%

\begin{figure}[t!]
     \centering
     \captionsetup{justification=centerlast, singlelinecheck=false}
     \includegraphics[width=1\linewidth]{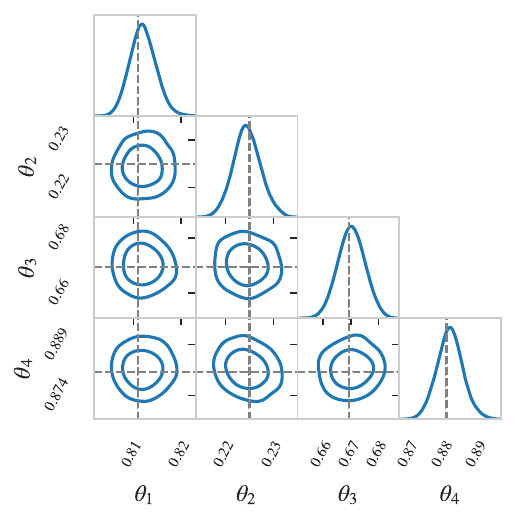}
     \caption{Neural Posterior Estimate of the four parameters for a single test event. The gray dashed lines indicate the true values while the blue contours represent the estimates at $68\%$ and $95\%$ confidence intervals.}
     \label{NPE_single}
 \end{figure}

The performance of the NPE for an individual randomly chosen test event is illustrated in Figure~\ref{NPE_single}. The diagonal panels show the one-dimensional marginal posterior distributions for each parameter, and the off-diagonal panels display the corresponding two-dimensional joint posteriors. The contours represent the $68\%$ and $95\%$ confidence intervals, enclosing the highest posterior density intervals under the assumption of a well-behaved posterior distribution. The vertical and horizontal dashed lines denote the corresponding true parameter values used to generate the plotted test data point. Overall, the posteriors are centered near the target values within the $68\%$ interval. The shape and orientation of the contours provide insight into the potential correlation among the parameters (elongated and tilted contours indicate correlations, while circular contours suggest independence). We do not see any correlation in our inferred posteriors, which is also evident on this randomly chosen test event.

\begin{figure}[t!]
     \centering
     \captionsetup{justification=centerlast, singlelinecheck=false}
     \includegraphics[width=1\linewidth]{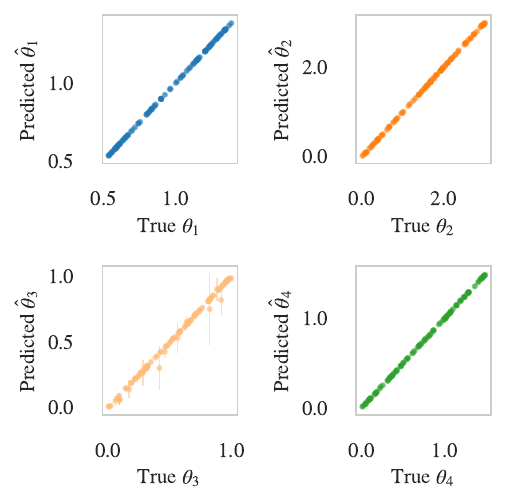}
     \caption{Comparison of true and predicted values for each parameter is shown for a subset of 200 test events, including $1\sigma$ error bars. Overall, the predictions closely follow the true values with narrow uncertainty bands, with only $\theta_3$ exhibiting some scatter and larger visible error bars.}
     \label{true_vs_pred}
 \end{figure}

To assess the overall performance of the model, on Figure~\ref{true_vs_pred} we show the comparison between the true $\theta^\mathrm{true}_i$ and predicted best fit values for each parameter $\hat{\theta}_i$ (where $i = 1,2,3,4$), on a sample of 200 events randomly chosen from our test set. Each point corresponds to an individual event prediction, with its associated $1\sigma$ error bars for each parameter (too small to be visible on the plot for all parameters except $\theta_3$), shown in different colors. All parameters are accurately reconstructed and lie close to the true values, with $\theta_3$ exhibiting somewhat larger deviations; nevertheless, in all cases, the true value remains within the predicted $1\sigma$ uncertainties.

\begin{figure}[t!]
     \centering
     \captionsetup{justification=centerlast, singlelinecheck=false}
     \includegraphics[width=1\linewidth]{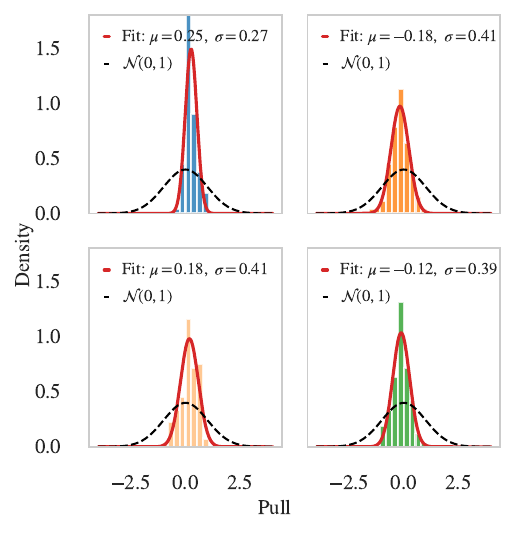}
     \caption{Histograms of pulls for 1000 test events for each parameter. All four parameter distributions (red solid line) are centered close to zero and somewhat narrower than the standard normal distribution (black dashed line), suggesting that the uncertainties may be overestimated.}
     \label{pulls}
 \end{figure}

We also utilize pull distributions to more closely examine how the NPE predicted uncertainties behave. In Figure~\ref{pulls} we present the pull distributions for each parameter, computed using the full test set of 1000 events. The pulls are defined as
\begin{equation}
    \textit{pull}_{j} = \frac{\hat{\theta}_{i,j} - \theta^{\mathrm{true}}_{i,j}}{\sigma_{i,j}},
\end{equation}
where $\sigma_{i,j}$ is the predicted NPE one-sigma error for each element of the test set, $j=0...1000$, and $i = 1,2,3,4$ corresponds to each of the four GENIE parameters. The color bars are the histograms for the pull for each parameter, the red solid curves represent Gaussian fits to the pulls, with the extracted mean
$\mu$ and width $\sigma$ indicated in the legend of each subplot, while the black dashed curves show the standard normal distribution
$N(0,1)$ expected for an unbiased and well-calibrated inference. We observe that all four distributions are centered close to zero but somewhat narrower than the standard distribution width, suggesting that the uncertainties may be overestimated.

\begin{figure}[t!]
     \centering
     \captionsetup{justification=centerlast, singlelinecheck=false}
     \includegraphics[width=1\linewidth]{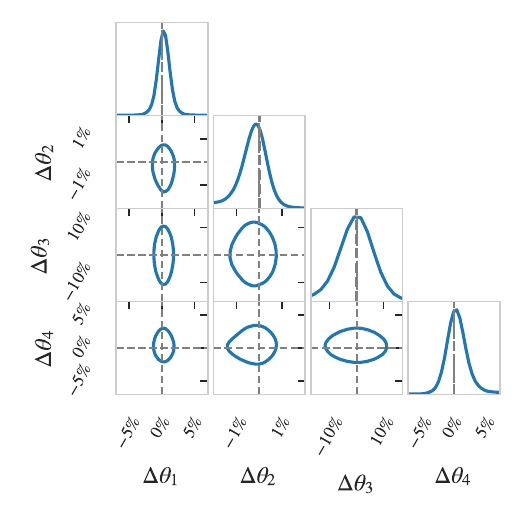}
     \caption{Residual distribution (in percentages) of four parameters for 1000 test events. The gray dashed lines indicate the zero value (perfect prediction). The distributions of residuals are very narrow and show no visible biases, i.e, they are all centered around zero. Additionally, all deviations are within a few percent.}
     \label{residuals}
 \end{figure}

To further assess the overall performance of the model, in Figure~\ref{residuals} we show the distribution of the residuals (in percentages) between model's best-fit estimate $\hat{\theta_i}$, and true parameter value $\theta_i^\mathrm{true}$ i.e., $\Delta\theta_i = (\hat{\theta_i} - \theta_i^\mathrm{true})/\theta_i^\mathrm{true}$ for $i=1,2,3,4$, computed over the same sample of $1000$ independent test events. Each diagonal panel displays the one-dimensional distribution of residuals for an individual parameter, while the off-diagonal panels illustrate the joint distributions between pairs of residuals. The dashed lines indicate zero bias. We observe that all residuals are centered around zero with narrow widths (up to a few percent, with only $\Delta\theta_3$ extending up to around $10\%$), indicating that the model yields unbiased estimates with low variance. 

  \begin{figure}[t!]
     \centering
     \captionsetup{justification=centerlast, singlelinecheck=false}
     \includegraphics[width=0.9\linewidth]{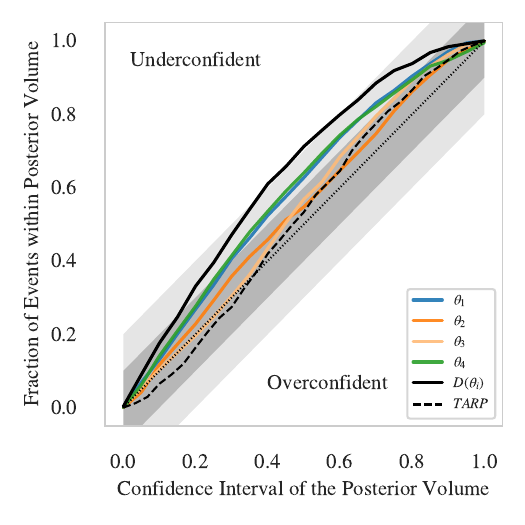}
     \caption{Posterior coverage of the $\theta_i$ parameters for $1000$ test events (colored solid lines). $D(\theta_i)$ and TARP curves show combined model performance over all parameters, and are plotted with solid and dashed black lines, respectively. The diagonal black dotted line indicates perfect uncertainty calibration. The gray regions indicate thresholds of 10\% (dark gray) and 20\% (light gray) uncertainty miscalibration. All curves are within the 20\% band and are showing slight underconfidence, which is preferred compared to an overconfident model with too narrow error bars.}
     \label{posterior_coverage}
 \end{figure}

In Figure~\ref{posterior_coverage}, we present the posterior coverage for each parameter $\theta_i$, evaluated again over the same 1000 test events. The solid colored lines represent the empirical coverage for each parameter, indicating the fraction of events where the true value lies within a given posterior confidence interval. The black solid curve represents the combined performance of the model on all parameters, for which we utilize a distance metric $D(\theta_i)$ from~\cite{WC2020,Poh2025}. $D(\theta_i)$ combines parameter values from posterior samples into a single objective function that takes into account the covariance between
different $\theta_i$ parameters. The black dashed line shows another common global diagnostic called the Truncated Acceptance Ratio Posterior (TARP)~\cite{lemos2023samplingbasedaccuracytestingposterior}. Given the posterior samples predicted by the model, the TARP diagnostic compares them to the corresponding ground-truth parameters by evaluating whether the true values fall within posterior credible regions at different confidence intervals. For each level, the method computes the fraction of true parameters contained within the inferred regions. We use the TARP implementation provided in the \texttt{sbi} diagnostics framework~\cite{lemos2023sampling}.
The black dotted diagonal corresponds to perfect uncertainty calibration, where the predicted confidence intervals exactly match the empirical coverage. The shaded bands around the diagonal (dark and light gray) denote 10\% and 20\% tolerance regions, respectively, for quantifying miscalibration. All of our inferred parameters are within the 10\%, with $\theta_1$ and $\theta_4$ leaking to the 20\% tolerance band for high posterior volumes. Note that all our parameters are underconfident, which is generally preferable to overconfidence, and it allows room for further improvement. More importantly, overconfidence would indicate unwarranted certainty in potentially incorrect results, thereby reducing the reliability of the model’s predictions. These conclusions are in line with the conclusions from the pulls (Figure~\ref{pulls}), which also show that the predicted errors are wider than what is necessary. Furthermore, even though the predicted posteriors of the individual parameter are wider than they need to be, this is not actually a problem for the analysis, as the largest uncertainties in the measurement come from the experimental errors (more details in Section~\ref{results}).

%%%%%%%%%%%%%%%%%%%%%%%%%%%%%%%%%%%%%%%%%
\section{Method validation and parameter measurement in data}
\label{results}
%%%%%%%%%%%%%%%%%%%%%%%%%%%%%%%%%%%%%%%%%

Our goal is to apply our SBI algorithm to determine optimal values of the four
GENIE simulation parameters originally tuned by
MicroBooNE~\cite{microboonegenietune}. These parameters control portions of
GENIE's cross-section calculation for nucleon knockout via the CCQE
($\theta_1$, $\theta_4$) and CCMEC ($\theta_2$, $\theta_3$)
channels. Their best-fit values were determined using a double-differential
measurement from T2K of charged-current interactions of muon neutrinos without
pions in the final state~\cite{T2K:2016jor}.\footnote{Two cross-section
measurements are reported in Ref.~\cite{T2K:2016jor}. Following MicroBooNE, we
use data from ``Analysis I'' for this study.} This T2K result is based on data
collected by the ND280 near detector, which used a plastic scintillator target
and was exposed to a beam of neutrinos with an average energy of 0.6 GeV at a
direction offset by 2.5$^{\circ}$ from the beam axis.

\begin{figure}
     \centering
     \captionsetup{justification=centerlast, singlelinecheck=false}
     \includegraphics[width=1\linewidth]{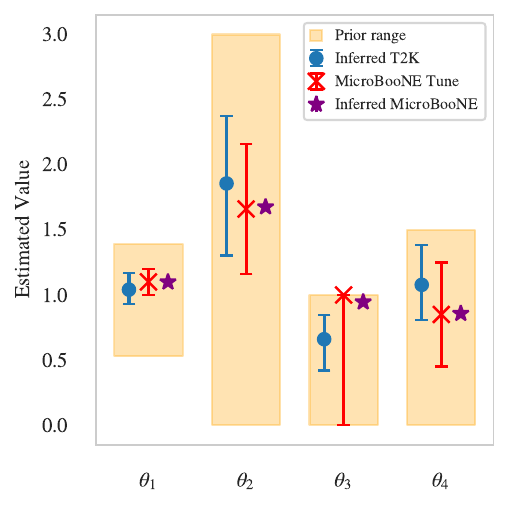}
     \caption{The red points correspond to the MicroBooNE fit parameters and uncertainties reported in Ref.~\cite{microboonegenietune}. In blue, we display the four parameters $\theta_i$ inferred by our network using T2K data as input, together with their corresponding error bars, computed using throws from the data covariance matrix as described on the main text. The orange bands indicate the prior ranges employed to train the SBI, as defined in Section~\ref{data}. As a validation, the violet points show the parameters inferred using the exact MicroBooNE histogram, exhibiting excellent agreement (with very small error bars, which are not visible on the figure) and thus corroborating the validity of our analysis. Overall, we observe a very good correspondence between the T2K-inferred parameters and the true MicroBooNE values.}
     \label{MicroBoonE_tune}
 \end{figure}

Before performing the actual data fit, we carried out two validation steps.
First, we verified that our NPE model is able to reproduce the correct
parameters when presented with the nominal histogram obtained from
the MicroBooNE Tune configuration of GENIE.
This test is equivalent to what is shown in
Fig.~\ref{NPE_single}, but choosing a configuration that was obtained by fitting
our target data set from T2K. As expected, our NPE model reproduces the MicroBooNE Tune parameters almost perfectly, which can be seen in the first two
rows of Table~\ref{table:results}. To compare these numbers, we also plot the original MicroBooNE Tune
values from~\cite{microboonegenietune} and the NPE inferred MicroBooNE values
in Figure~\ref{MicroBoonE_tune}, as red and violet points respectively.

\begin{table*}[hbt!]
\centering
\captionsetup{justification=centerlast, singlelinecheck=false}
\begin{threeparttable}
\scalebox{0.925}{
\begin{tabular}{c | c c c c | c c c c | c}
 Model & $\theta_1$ & $\theta_2$ & $\theta_3$ & $\theta_4$ & $\sigma_{\theta_1}$ & $\sigma_{\theta_2}$ & $\sigma_{\theta_3}$ & $\sigma_{\theta_4}$ & $\chi^2$(T2K data, 58 bins)\tnote{c} \\ [0.5ex]
 \hline\hline
 MicroBooNE Tune\tnote{a} &1.10 & 1.66 & 1.00 & 0.85 &$\pm 0.10$ & $\pm 0.50$ &+1.00(-0.00) & $\pm 0.40$& 115.10 \\ [0.5ex]
 Inferred MicroBooNE Tune\tnote{b}\; & 1.09 & 1.67 & 0.94 & 0.85 & $\pm 0.003$ & $\pm 0.01$ & $\pm 0.01$ & $\pm 0.007$ & 113.07 \\ [0.5ex]
 \hline
 NuWro & - & - & - & - & - & - & - & - & 126.73 \\ [0.5ex]
 Inferred NuWro\tnote{b}  & 1.10 & 0.70 & 0.09 & 0.57 &$\pm 0.003$ & $\pm 0.015$ & $\pm 0.013$ & $\pm 0.007$& 126.93 \\ [0.5ex]
 \hline
 Inferred T2K\tnote{a}  & 1.04 & 1.85 & 0.65 & 1.07 & +0.11(-0.12) & +0.55(-0.24) & +0.24(-0.18)& +0.27(-0.30) & 107.29 \\ [0.5ex]
\end{tabular}
}

\caption{Inferred parameters, their uncertainties, and the corresponding $\chi^2$ values obtained when
comparing with the T2K measurement. Some values are also presented in Fig.~\ref{MicroBoonE_tune}: MicroBooNE Tune (red), Inferred MicroBooNE Tune (violet), and Inferred T2K (blue).
The agreement between the original and NPE fits is excellent, both for the MicroBooNE Tune and NuWro. The NPE fit to the T2K data provides the overall best $\chi^2$. Not all $\theta$ parameters are shared between the NuWro and GENIE physics models, so we omit their values in the NuWro row of the table.}\label{table:results}
\end{threeparttable}
\begin{tablenotes}
  \item[a] ${}^a$ The $\sigma$ errors correspond to experimental errors.
  \item[b] ${}^b$ The $\sigma$ errors correspond to one standard deviation of the NPE predicted posteriors.
  \item[c] ${}^c$ $\chi^2$ values reported here include all measurement bins shown in Figs.~\ref{fig:nuwro_main}~and~\ref{fig:nuwro_all}. The full data covariance matrix (including off-diagonal elements) is used in the $\chi^2$ calculation.
  \end{tablenotes}
\end{table*}

\begin{figure*}[hbt!]
\centering

\begin{subfigure}{0.45\textwidth}
    \centering
    \includegraphics[width=\linewidth]{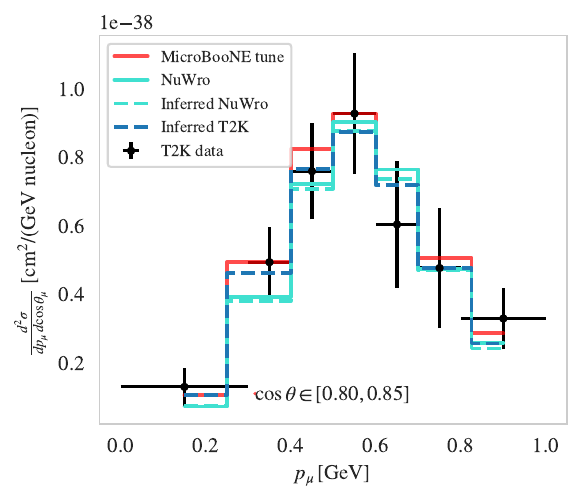}
    %\caption{NuWro slice 4}
    \label{fig:nuwro2}
\end{subfigure}
\hfill
\begin{subfigure}{0.45\textwidth}
    \centering
    \includegraphics[width=\linewidth]{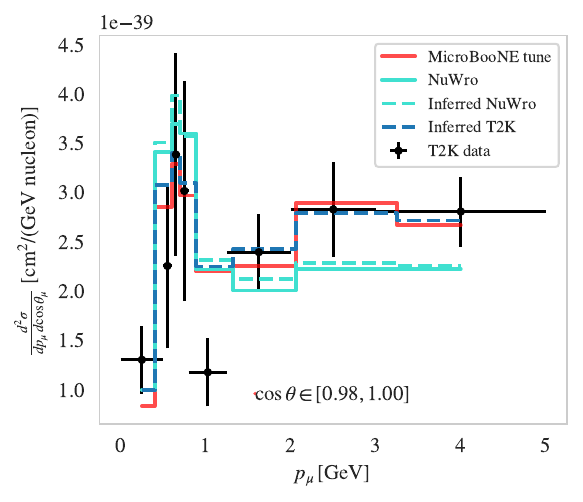}
    %\caption{NuWro slice 8}
    \label{fig:nuwro3}
\end{subfigure}

\vspace{0.2cm}

\caption{Comparison of the T2K cross-section data (black points) to the NuWro~\cite{Juszczak:2009qa} (solid cyan), NPE-inferred NuWro (dashed cyan), MicroBooNE Tune~\cite{microboonegenietune} (solid red), and NPE-inferred T2K (dashed blue) model predictions in two different slices of $\cos\theta$.}
\label{fig:nuwro_main}
\end{figure*}

As a next step, we consider as input a cross-section prediction for the T2K
measurement calculated using the default configuration of version 25.03.2 of the NuWro event generator~\cite{Juszczak:2009qa}. While
the physics model adopted in this version of NuWro has some similarities with
GENIE, the details of each implementation differ sufficiently that an exact
match between the two simulations cannot be achieved solely by parameter
adjustments. Nevertheless, due to the limited capability of current experimental
workflows to support large-scale simulations with multiple neutrino event
generators, the prospect of a computationally cheap method for approximating
the predictions of one generator with another is worthy of study. To
investigate this possibility, we treated the NuWro prediction of the T2K
measurement as mock data. In addition to providing a test of how well SBI-based
modifications to our four GENIE parameters could perform as a surrogate model
for NuWro; this also served as an independent test of our tuning procedure
before confronting it with the measured T2K results. The values of the GENIE
model parameters inferred from an input cross-section histogram calculated with
NuWro are shown in the fourth row of Table~\ref{table:results}. Although the
inferred GENIE approximation to NuWro is not a perfect match, we note that the
overall $\chi^2$ metric for goodness-of-fit with the T2K data is nearly
identical between NuWro and its GENIE surrogate model. Bin-by-bin visual
agreement can be qualitatively assessed between the original NuWro prediction
and the ``Inferred NuWro'' surrogate model in Fig.~\ref{fig:nuwro_main}. Each plot in the figure shows the muon momentum distribution measured by T2K (black points) within a particular range of $\cos\theta$, the cosine of the angle between the incoming neutrino and outgoing muon. The two angular bins shown represent a portion of the full data set used in our study, with the remainder plotted in Appendix~\ref{appA}. The prediction of the original MicroBooNE Tune model is plotted as the solid red line. The NuWro and Inferred NuWro model
predictions are drawn as solid and dashed cyan lines, respectively. While the
quality of the agreement between these last two models varies across bins of
the measurement, in general, we find that the surrogate model mimics the
original NuWro prediction reasonably well. We attribute residual
differences between the original and inferred NuWro results to two sources.
First, the GENIE parameters used in the MicroBooNE Tune may imperfectly
represent relevant degrees of freedom considered in the NuWro model. In
particular, it seems unlikely that any allowed value of $\theta_3$ would be
able to fully reproduce the differential shape of the NuWro cross-section for
the two-nucleon knockout process. Second, there are known differences between
other aspects of the GENIE and NuWro physics models that are unaffected by the
parameters studied here. An important example for the T2K data set of interest
is the different approach taken by each code to intranuclear absorption of
pions produced in inelastic neutrino-nucleon interaction channels. Despite
these caveats, however, the level of agreement achieved here between the NuWro
and Inferred NuWro predictions may motivate a more sophisticated pursuit of
SBI-based surrogate models for neutrino event generators.

In Figure~\ref{fig:nuwro_main} and the accompanying figures in
Appendix~\ref{appA}, we also plot in dashed blue the cross-section prediction
obtained with our SBI algorithm using the actual T2K measurement as input. The
inferred parameter values and $\chi^2$ metric for this SBI-based GENIE
predictions are also shown in the final row of Table~\ref{table:results}. While the
dashed blue prediction shows the result of running inference directly on the
central-value data points reported by T2K, a full covariance matrix describing
the measurement uncertainties is also provided in the collaboration's data
release. In the original MicroBooNE Tune
publication~\cite{microboonegenietune}, the authors reported difficulty in
using the full covariance matrix within their tuning algorithm, which was based
on numerical minimization of $\chi^2$ in repeated comparisons to the data.
Ultimately, a common pathology of similar procedures, known as Peelle's
Pertinent Puzzle (PPP)~\cite{fruhwirth2012peelle}, led MicroBooNE to neglect
the off-diagonal covariance matrix elements in their final fit, effectively
ignoring the significant measurement correlations reported by T2K. While this
choice was cross-checked in the original analysis using a different technique
involving a ``norm-shape'' transformation of the covariance
matrix~\cite{Chakrani:2023htw, Koch:2021yda} (which yielded a compatible fit), MicroBooNE's approach has been shown to yield poor results when
applied to other data sets~\cite{Wolfs:2025ofb}.

\begin{figure}[t!]
     \centering
     \captionsetup{justification=centerlast, singlelinecheck=false}
     \includegraphics[width=1\linewidth]{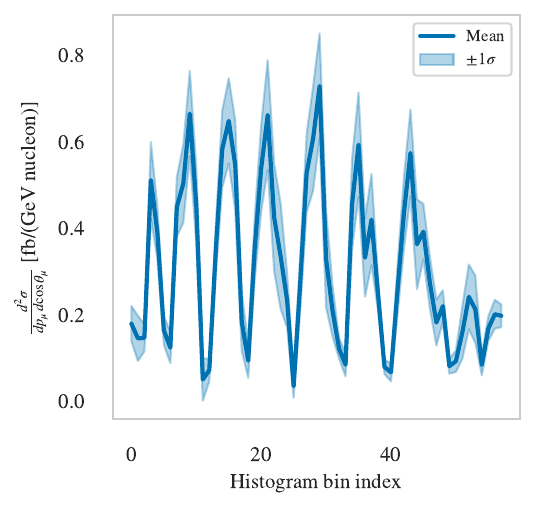}
     \caption{Variation of the T2K histograms over one thousand simulated events. The solid line denotes the mean (central) value, while the shaded blue region represents the corresponding $1\sigma$ variations obtained from samples drawn using the covariance matrix.}
     \label{hist_variation}
 \end{figure}

The general problem of PPP for interaction model tuning is widely recognized in
the neutrino community, and discussions continue on optimal methods for
addressing it~\cite{ppp}. A strength of our SBI strategy is that the issue is
sidestepped entirely; inference is performed each time using a single histogram
without associated uncertainties, so the problematic correlations do not enter
into the calculation. To propagate the measurement uncertainties into our results
while preserving correlations, we first generated an ensemble of one thousand
variations of the T2K measurement using random throws from a multivariate
zero-mean Gaussian distribution using the data covariance matrix.
Figure~\ref{hist_variation} shows the mean (solid line) and the corresponding
standard deviation (shaded band) obtained from this procedure. To plot the
entire two-dimensional measurement on a single axis, we show the cross-section
as a function of histogram bin index, which ranges from $0$ to $58$.

\begin{figure}[t!]
     \centering
     \captionsetup{justification=centerlast, singlelinecheck=false}
     \includegraphics[width=\linewidth]{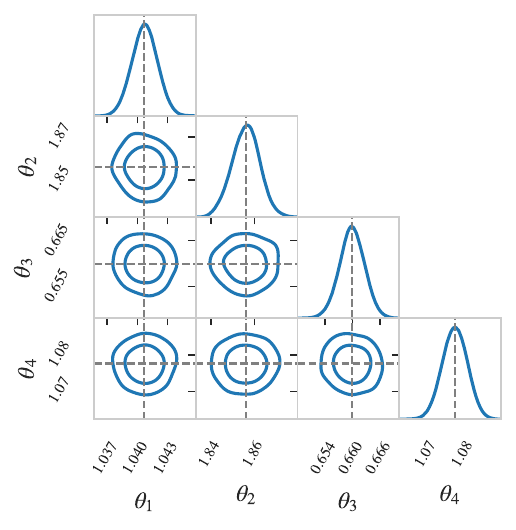}
     \caption{Neural Posterior Estimate of the four parameters for the T2K central value. The gray dashed lines represent the inferred central values while the blue contours represent the estimates at $68\%$ and $95\%$ confidence intervals.}
     \label{NPE_t2k0}
 \end{figure}

Figure~\ref{NPE_t2k0} shows the NPE predicted posteriors for the T2K central
value measurement. The extracted parameters are indicated by the gray dashed
lines, while the blue contours represent the $68\%$ and $95\%$ confidence
intervals from the NPE fit.

In Figure~\ref{MicroBoonE_tune}, the inferred parameter values obtained
using the T2K central values (same as in Figure~\ref{NPE_t2k0}) are shown as
blue dots. Here we plot the corresponding $1\sigma$ error bars, also in blue, which are
computed as follows. First, we compute the distribution of deviations between
each prediction obtained from the covariance matrix histogram throws,
$\hat{\theta}_\mathrm{T2K}^{i,j}$ and the central T2K value
$\hat{\theta}_\mathrm{T2K_0}$. For each sample $i$, we define \[ \Delta
\hat{\theta}_\mathrm{T2K}^{i,j} = \hat{\theta}_\mathrm{T2K}^{i,j} -
\hat{\theta}_\mathrm{T2K_0}^i, \] were again $j=0,...,1000$ and $i=1,2,3,4$.
Then, the full set of deviations is stacked into an array, from which the
median and the $16th$ and $84th$ percentiles are calculated along each
parameter axis. These percentiles are then used to define asymmetric 68\%
confidence intervals corresponding to the $1\sigma$ error bars. As expected,
these error bars are much wider than errors coming from a single predicted
posterior from the NPE model (for example, for the central T2K value on
Figure~\ref{NPE_t2k0}). Numerical values of the blue $1\sigma$ uncertainties are reported in the final row of Table~\ref{table:results}. Figure~\ref{MicroBoonE_tune} also shows that the
inferred values (blue points with error bars) are in close agreement with the
MicroBooNE Tune~\cite{microboonegenietune}, indicated in red, with all our
inferred values being within MicroBooNE error bars. We also show the prior
range (orange bands) for all $\theta_i$ parameters used to generate the
training data for our NPE model. Finally, the violet points show the parameters
inferred using the exact MicroBooNE histogram, which are in excellent agreement
with true MicroBooNE parameters.

In summary, in Table~\ref{table:results} we report the parameter values for all
of our tests and the corresponding $\chi^2$ values obtained when comparing with
the T2K measurement. The agreement between the original and NPE fits is
excellent, both for the MicroBooNE Tune and NuWro. The NPE fit to T2K data
provides the overall best $\chi^2$.

%%%%%%%%%%%%%%%%%%%%%%%%%%%%%%%%%%%%%%%%%
\section{Summary and Conclusion}
\label{summary}
%%%%%%%%%%%%%%%%%%%%%%%%%%%%%%%%%%%%%%%%%

Shortcomings in neutrino scattering models are commonly addressed by tuning simulation parameters to match interaction data, a process that is growing ever more demanding. For the first time, here we show that simulation-based inference with neural posterior estimation can meet this challenge by efficiently inferring neutrino interaction model parameters from data.

As a first application, we train a model that infers the four GENIE parameters that control the quasi-elastic and MEC processes in charged-current neutrino-nucleus interactions. We set up the model to replicate the procedure followed by MicroBooNE to extract those parameters from T2K data. After validating the model on different GENIE configurations, including the MicroBooNE Tune, we derived a GENIE configuration that approximates the predictions of the NuWro generator for the T2K data set. This effectively defines a procedure for deriving surrogate models of other generators, potentially allowing experiments to save computing resources and avoid producing a full simulation sample for each generator of interest. Finally, we inferred the parameters on the same T2K measurement used for the MicroBooNE Tune. We find that the parameters extracted by our model are within the uncertainties of the MicroBooNE result and provide a slightly improved fit to the data when evaluated in terms of $\chi^2$.

The method we have developed in this article can be further developed in future work. This may include understanding how the model scales to more challenging problems, both in terms of the number of parameters fitted and in terms of the number of bins included in the data. Future studies may also focus on uncertainty quantification, improving the model coverage, and evaluating strategies for including experimental uncertainties as part of the model inputs.
Overall, we expect that our approach will be generalizable to similar problems in neutrino interaction modeling, potentially leading to more powerful parameter fits that are currently limited by personnel effort and computing resources.

%%%%%%%%%%%%%%%%%%%%%%%%%%%%%%%%%%%%

\section{Acknowledgments}
This work was produced by Fermi Forward Discovery Group, LLC under Contract No. 89243024CSC000002 with the U.S. Department of Energy, Office of Science, Office of High Energy Physics. The United States Government retains, and the publisher, by accepting the work for publication, acknowledges that the United States Government retains a non-exclusive, paid-up, irrevocable, world-wide license to publish or reproduce the published form of this work, or allow others to do so, for United States Government purposes. The Department of Energy will provide public access to these results of federally sponsored research in accordance with the DOE Public Access Plan
(\url{http://energy.gov/downloads/doe-public-access-plan}).
The work of K.T. is supported by DOE Grants DE-SCL0000156 and FNAL 23-32.

%ECA internal fnal number: KA2401045

\textbf{Author Contributions:}

\noindent Tame-Narvaez: Conceptualization, Methodology, Investigation, Formal analysis, Software, Validation, Data curation, Resources, Visualization, Writing - Original Draft, Writing - Review \& Editing.\\
\noindent Gardiner: Conceptualization, Methodology, Formal analysis, Software, Validation, Data curation, Resources, Visualization, Writing - Original Draft, Writing - Review \& Editing, Supervision.\\
\noindent \'{C}iprijanovi\'{c}: Conceptualization, Methodology, Writing - Original Draft, Writing - Review \& Editing, Supervision.\\
\noindent Cerati:  Conceptualization, Methodology, Writing - Original Draft, Writing - Review \& Editing, Supervision.\\

%\section*{References}
\bibliography{sbi}

@ARTICLE{CB2020,
       author = {{Cranmer}, Kyle and {Brehmer}, Johann and {Louppe}, Gilles},
        title = "{The frontier of simulation-based inference}",
      journal = {Proceedings of the National Academy of Science},
         year = 2020,
        month = dec,
       volume = {117},
       number = {48},
        pages = {30055-30062},
          doi = {10.1073/pnas.1912789117},
archivePrefix = {arXiv},
       eprint = {1911.01429},
 primaryClass = {stat.ML},
       adsurl = {https://ui.adsabs.harvard.edu/abs/2020PNAS..11730055C}
}

@article{Rubin1984,
 ISSN = {00905364, 21688966},
 URL = {http://www.jstor.org/stable/2240995},
 author = {Donald B. Rubin},
 journal = {The Annals of Statistics},
 number = {4},
 pages = {1151--1172},
 publisher = {Institute of Mathematical Statistics},
 title = {Bayesianly Justifiable and Relevant Frequency Calculations for the Applied Statistician},
 urldate = {2023-06-27},
 volume = {12},
 year = {1984}
}

@article{brehmer2018guide,
   title={A guide to constraining effective field theories with machine learning},
   volume={98},
   ISSN={2470-0029},
   url={http://dx.doi.org/10.1103/PhysRevD.98.052004},
   DOI={10.1103/physrevd.98.052004},
   number={5},
   journal={\prd},
   publisher={American Physical Society (APS)},
   author={Brehmer, Johann and Cranmer, Kyle and Louppe, Gilles and Pavez, Juan},
   year={2018},
   month=sep }

@ARTICLE{Poh2025,
       author = {{Poh}, Jason and {Samudre}, Ashwin and {{\'C}iprijanovi{\'c}}, Aleksandra and {Frieman}, Joshua and {Khullar}, Gourav and {Nord}, Brian D.},
        title = "{Deep inference of simulated strong lenses in ground-based surveys}",
      journal = {J. Cosmol. Astropart. Phys.},
         year = 2025,
        month = may,
       volume = {2025},
       number = {5},
          eid = {053},
        pages = {053},
          doi = {10.1088/1475-7516/2025/05/053},
archivePrefix = {arXiv},
       eprint = {2501.08524},
 primaryClass = {astro-ph.IM},
       adsurl = {https://ui.adsabs.harvard.edu/abs/2025JCAP...05..053P}
}

@ARTICLE{Khullar2022,
       author = {{Khullar}, Gourav and {Nord}, Brian and {{\'C}iprijanovi{\'c}}, Aleksandra and {Poh}, Jason and {Xu}, Fei},
        title = "{DIGS: deep inference of galaxy spectra with neural posterior estimation}",
      journal = {Machine Learning: Science and Technology},
         year = 2022,
        month = dec,
       volume = {3},
       number = {4},
          eid = {04LT04},
        pages = {04LT04},
          doi = {10.1088/2632-2153/ac98f4},
archivePrefix = {arXiv},
       eprint = {2211.09126},
 primaryClass = {astro-ph.GA},
       adsurl = {https://ui.adsabs.harvard.edu/abs/2022MLS&T...3dLT04K}
}

@INPROCEEDINGS{Reza2022,
       author = {{Reza}, Moonzarin and {Zhang}, Yuanyuan and {Nord}, Brian and {Poh}, Jason and {Ciprijanovic}, Aleksandra and {Strigari}, Louis},
        title = "{Estimating Cosmological Constraints from Galaxy Cluster Abundance using Simulation-Based Inference}",
    booktitle = {Machine Learning for Astrophysics},
         year = 2022,
        month = jul,
          eid = {20},
        pages = {20},
          doi = {10.48550/arXiv.2208.00134},
archivePrefix = {arXiv},
       eprint = {2208.00134},
 primaryClass = {astro-ph.CO},
       adsurl = {https://ui.adsabs.harvard.edu/abs/2022mla..confE..20R}
}

@ARTICLE{GF2021,
       author = {{Gerardi}, Francesca and {Feeney}, Stephen M. and {Alsing}, Justin},
        title = "{Unbiased likelihood-free inference of the Hubble constant from light standard sirens}",
      journal = {\prd},
         year = 2021,
        month = oct,
       volume = {104},
       number = {8},
          eid = {083531},
        pages = {083531},
          doi = {10.1103/PhysRevD.104.083531},
archivePrefix = {arXiv},
       eprint = {2104.02728},
 primaryClass = {astro-ph.CO},
       adsurl = {https://ui.adsabs.harvard.edu/abs/2021PhRvD.104h3531G}
}

@article{Brehmer_2019,
	doi = {10.3847/1538-4357/ab4c41},

	url = {https://doi.org/10.3847%2F1538-4357%2Fab4c41},

	year = 2019,
	month = {nov},

	publisher = {American Astronomical Society},

	volume = {886},

	number = {1},

	pages = {49},

	author = {Johann Brehmer and Siddharth Mishra-Sharma and Joeri Hermans and Gilles Louppe and Kyle Cranmer},

	title = {Mining for Dark Matter Substructure: Inferring Subhalo Population Properties from Strong Lenses with Machine Learning},

	journal = {\apj}
}

@ARTICLE{Coogan2022,
       author = {{Coogan}, Adam and {Anau Montel}, Noemi and {Karchev}, Konstantin and {Grootes}, Meiert W. and {Nattino}, Francesco and {Weniger}, Christoph},
        title = "{One never walks alone: the effect of the perturber population on subhalo measurements in strong gravitational lenses}",
      journal = {arXiv e-prints},
         year = 2022,
        month = sep,
          eid = {arXiv:2209.09918},
        pages = {arXiv:2209.09918},
archivePrefix = {arXiv},
       eprint = {2209.09918},
 primaryClass = {astro-ph.CO},
       adsurl = {https://ui.adsabs.harvard.edu/abs/2022arXiv220909918C}
}

@ARTICLE{Anau2023,
       author = {{Anau Montel}, Noemi and {Coogan}, Adam and {Correa}, Camila and {Karchev}, Konstantin and {Weniger}, Christoph},
        title = "{Estimating the warm dark matter mass from strong lensing images with truncated marginal neural ratio estimation}",
      journal = {Mon. Not. R. Astron. Soc.},
         year = 2023,
        month = jan,
       volume = {518},
       number = {2},
        pages = {2746-2760},
          doi = {10.1093/mnras/stac3215},
archivePrefix = {arXiv},
       eprint = {2205.09126},
 primaryClass = {astro-ph.CO},
       adsurl = {https://ui.adsabs.harvard.edu/abs/2023MNRAS.518.2746A}
}

@ARTICLE{Legin2021,
       author = {{Legin}, Ronan and {Hezaveh}, Yashar and {Perreault Levasseur}, Laurence and {Wandelt}, Benjamin},
        title = "{Simulation-Based Inference of Strong Gravitational Lensing Parameters}",
      journal = {arXiv e-prints},
         year = 2021,
        month = dec,
          eid = {arXiv:2112.05278},
        pages = {arXiv:2112.05278},
archivePrefix = {arXiv},
       eprint = {2112.05278},
 primaryClass = {astro-ph.CO},
       adsurl = {https://ui.adsabs.harvard.edu/abs/2021arXiv211205278L}
}

@ARTICLE{WagnerCarena2022,
       author = {{Wagner-Carena}, Sebastian and {Aalbers}, Jelle and {Birrer}, Simon and {Nadler}, Ethan O. and {Darragh-Ford}, Elise and {Marshall}, Philip J. and {Wechsler}, Risa H.},
        title = "{From Images to Dark Matter: End-to-end Inference of Substructure from Hundreds of Strong Gravitational Lenses}",
      journal = {\apj},
         year = 2023,
        month = jan,
       volume = {942},
       number = {2},
          eid = {75},
        pages = {75},
          doi = {10.3847/1538-4357/aca525},
archivePrefix = {arXiv},
       eprint = {2203.00690},
 primaryClass = {astro-ph.CO},
       adsurl = {https://ui.adsabs.harvard.edu/abs/2023ApJ...942...75W}
}

@misc{papamakarios2018maskedautoregressiveflowdensity,
      title={Masked Autoregressive Flow for Density Estimation},
      author={George Papamakarios and Theo Pavlakou and Iain Murray},
      year={2018},
      eprint={1705.07057},
      archivePrefix={arXiv},
      primaryClass={stat.ML},
      url={https://arxiv.org/abs/1705.07057},
}

@ARTICLE{BR2021,
       author = {{Brehmer}, Johann},
        title = "{Simulation-based inference in particle physics}",
      journal = {Nature Reviews Physics},
         year = 2021,
        month = may,
       volume = {3},
       number = {5},
        pages = {305-305},
          doi = {10.1038/s42254-021-00305-6},
archivePrefix = {arXiv},
       eprint = {2010.06439},
 primaryClass = {hep-ph},
       adsurl = {https://ui.adsabs.harvard.edu/abs/2021NatRP...3..305B}
}

@ARTICLE{MN2024,
       author = {{Mastandrea}, Radha and {Nachman}, Benjamin and {Plehn}, Tilman},
        title = "{Constraining the Higgs potential with neural simulation-based inference for di-Higgs production}",
      journal = {\prd},
         year = 2024,
        month = sep,
       volume = {110},
       number = {5},
          eid = {056004},
        pages = {056004},
          doi = {10.1103/PhysRevD.110.056004},
archivePrefix = {arXiv},
       eprint = {2405.15847},
 primaryClass = {hep-ph},
       adsurl = {https://ui.adsabs.harvard.edu/abs/2024PhRvD.110e6004M}
}

@ARTICLE{BM2024,
       author = {{Barru{\'e}}, Ricardo and {Mu{\'\i}{\~n}o}, Patricia Conde and {Dao}, Valerio and {Santos}, Rui},
        title = "{Simulation-based inference in the search for CP violation in leptonic WH production}",
      journal = {Journal of High Energy Physics},
         year = 2024,
        month = apr,
       volume = {2024},
       number = {4},
          eid = {14},
        pages = {14},
          doi = {10.1007/JHEP04(2024)014},
archivePrefix = {arXiv},
       eprint = {2308.02882},
 primaryClass = {hep-ph},
       adsurl = {https://ui.adsabs.harvard.edu/abs/2024JHEP...04..014B}
}

@ARTICLE{BB2021,
       author = {{Bieringer}, Sebastian and {Butter}, Anja and {Heimel}, Theo and {H{\"o}che}, Stefan and {K{\"o}the}, Ullrich and {Plehn}, Tilman and {Radev}, Stefan T.},
        title = "{Measuring QCD Splittings with Invertible Networks}",
      journal = {SciPost Physics},
         year = 2021,
        month = jun,
       volume = {10},
       number = {6},
          eid = {126},
        pages = {126},
          doi = {10.21468/SciPostPhys.10.6.126},
archivePrefix = {arXiv},
       eprint = {2012.09873},
 primaryClass = {hep-ph},
       adsurl = {https://ui.adsabs.harvard.edu/abs/2021ScPP...10..126B}
}

@Article{	  genie2010,
  title		= {{The GENIE neutrino Monte Carlo generator}},
  journal	= {Nucl. Instrum. Methods Phys. Res. A},
  volume	= {614},
  number	= {1},
  pages		= {87--104},
  year		= {2010},
  issn		= {0168-9002},
  doi		= {10.1016/j.nima.2009.12.009},
  author	= {C. Andreopoulos and others},
  eprint	= {0905.2517},
  primaryclass	= {hep-ph},
  archiveprefix	= {arXiv}
}

@article{Gavrikov:2025rps,
    author = "Gavrikov, Arsenii and others",
    title = "{Simulation-based inference for precision neutrino physics through neural Monte Carlo tuning}",
    eprint = "2507.23297",
    archivePrefix = "arXiv",
    primaryClass = "physics.data-an",
    doi = "10.1038/s42005-026-02499-6",
    journal = "Commun. Phys.",
    volume = "9",
    number = "1",
    pages = "63",
    year = "2026"
}

@Article{	  microboonegenietune,
  title		= {New {$\mathrm{CC}0\ensuremath{\pi}$ GENIE} model tune for
		  {MicroBooNE}},
  author	= {Abratenko, P. and others},
  collaboration	= {{MicroBooNE collaboration}},
  journal	= {Phys. Rev. D},
  volume	= {105},
  issue		= {7},
  pages		= {072001},
  numpages	= {23},
  year		= {2022},
  month		= {Apr},
  publisher	= {American Physical Society},
  doi		= {10.1103/PhysRevD.105.072001},
  eprint	= {2110.14028},
  archiveprefix	= {arXiv},
  primaryclass	= {hep-ex}
}

@article{nuisance,
  author = "Stowell, P. and others",
  title = "{NUISANCE: a neutrino cross-section generator tuning and comparison framework}",
  eprint = "1612.07393",
  archivePrefix = "arXiv",
  primaryClass = "hep-ex",
  doi = "10.1088/1748-0221/12/01/P01016",
  journal = "JINST",
  volume = "12",
  number = "01",
  pages = "P01016",
  year = "2017"
}

@Article{	  genie2021,
  author	= "Alvarez-Ruso, Luis and others",
  collaboration	= "{GENIE} collaboration",
  title		= "{Recent highlights from GENIE v3}",
  eprint	= "2106.09381",
  archiveprefix	= "arXiv",
  primaryclass	= "hep-ph",
  reportnumber	= "FERMILAB-PUB-21-266-SCD-T",
  doi		= "10.1140/epjs/s11734-021-00295-7",
  journal	= "Eur. Phys. J. ST",
  volume	= "230",
  number	= "24",
  pages		= "4449--4467",
  year		= "2021"
}

@article{GENIE:2024ufm,
    author = "Li, Weijun and others",
    collaboration = "GENIE",
    title = "{First combined tuning on transverse kinematic imbalance data with and without pion production constraints}",
    eprint = "2404.08510",
    archivePrefix = "arXiv",
    primaryClass = "hep-ex",
    reportNumber = "FERMILAB-PUB-24-0122-CSAID-PPD",
    doi = "10.1103/PhysRevD.110.072016",
    journal = "Phys. Rev. D",
    volume = "110",
    number = "7",
    pages = "072016",
    year = "2024"
}

@article{GENIE:2022qrc,
    author = "Tena-Vidal, Julia and others",
    collaboration = "GENIE",
    title = "{Neutrino-nucleus CC0$\pi$ cross-section tuning in GENIE v3}",
    eprint = "2206.11050",
    archivePrefix = "arXiv",
    primaryClass = "hep-ph",
    reportNumber = "FERMILAB-PUB-22-296-ND-QIS-SCD",
    doi = "10.1103/PhysRevD.106.112001",
    journal = "Phys. Rev. D",
    volume = "106",
    number = "11",
    pages = "112001",
    year = "2022"
}

@article{Coyle:2022bwa,
    author = "Coyle, Nina M. and Li, Shirley Weishi and Machado, Pedro A. N.",
    title = "{The impact of neutrino-nucleus interaction modeling on new physics searches}",
    eprint = "2210.03753",
    archivePrefix = "arXiv",
    primaryClass = "hep-ph",
    reportNumber = "FERMILAB-PUB-22-726-T",
    doi = "10.1007/JHEP12(2022)166",
    journal = "JHEP",
    volume = "12",
    pages = "166",
    year = "2022"
}

@article{GENIE:2021wox,
    author = "Tena-Vidal, J{\'u}lia and others",
    collaboration = "GENIE",
    title = "{Hadronization model tuning in genie v3}",
    eprint = "2106.05884",
    archivePrefix = "arXiv",
    primaryClass = "hep-ph",
    reportNumber = "FERMILAB-PUB-21-024-QIS-SCD-T",
    doi = "10.1103/PhysRevD.105.012009",
    journal = "Phys. Rev. D",
    volume = "105",
    number = "1",
    pages = "012009",
    year = "2022"
}

@article{GENIE:2021zuu,
    author = "Tena-Vidal, J{\'u}lia and others",
    collaboration = "GENIE",
    title = "{Neutrino-nucleon cross-section model tuning in GENIE v3}",
    eprint = "2104.09179",
    archivePrefix = "arXiv",
    primaryClass = "hep-ph",
    reportNumber = "FERMILAB-PUB-20-531-SCD-T",
    doi = "10.1103/PhysRevD.104.072009",
    journal = "Phys. Rev. D",
    volume = "104",
    number = "7",
    pages = "072009",
    year = "2021"
}

@article{MINERvA:2019kfr,
    author = "Stowell, P. and others",
    collaboration = "MINERvA",
    title = "{Tuning the GENIE Pion Production Model with MINER$\nu$A Data}",
    eprint = "1903.01558",
    archivePrefix = "arXiv",
    primaryClass = "hep-ex",
    reportNumber = "FERMILAB-PUB-19-093-ND",
    doi = "10.1103/PhysRevD.100.072005",
    journal = "Phys. Rev. D",
    volume = "100",
    number = "7",
    pages = "072005",
    year = "2019"
}

@article{NOvA:2020rbg,
    author = "Acero, M. A. and others",
    collaboration = "NOvA, R. Group",
    title = "{Adjusting neutrino interaction models and evaluating uncertainties using NOvA near detector data}",
    eprint = "2006.08727",
    archivePrefix = "arXiv",
    primaryClass = "hep-ex",
    reportNumber = "FERMILAB-PUB-20-243-ND",
    doi = "10.1140/epjc/s10052-020-08577-5",
    journal = "Eur. Phys. J. C",
    volume = "80",
    number = "12",
    pages = "1119",
    year = "2020"
}

@article{T2K:2016jor,
    author = "Abe, Ko and others",
    collaboration = "T2K",
    title = "{Measurement of double-differential muon neutrino charged-current interactions on C$_8$H$_8$ without pions in the final state using the T2K off-axis beam}",
    eprint = "1602.03652",
    archivePrefix = "arXiv",
    primaryClass = "hep-ex",
    doi = "10.1103/PhysRevD.93.112012",
    journal = "Phys. Rev. D",
    volume = "93",
    number = "11",
    pages = "112012",
    year = "2016"
}

@InProceedings{lemos2023sampling,
  title = 	 {Sampling-Based Accuracy Testing of Posterior Estimators for General Inference},
  author =       {Lemos, Pablo and Coogan, Adam and Hezaveh, Yashar and Perreault-Levasseur, Laurence},
  booktitle = 	 {Proceedings of the 40th International Conference on Machine Learning},
  pages = 	 {19256--19273},
  year = 	 {2023},
  editor = 	 {Krause, Andreas and Brunskill, Emma and Cho, Kyunghyun and Engelhardt, Barbara and Sabato, Sivan and Scarlett, Jonathan},
  volume = 	 {202},
  series = 	 {Proceedings of Machine Learning Research},
  month = 	 {23--29 Jul},
  publisher =    {PMLR},
  url = 	 {https://proceedings.mlr.press/v202/lemos23a.html}
}

@misc{DB2025,
  title = {Simulation-{{Based Inference}}: {{A Practical Guide}}},
  author = {Deistler, Michael and others},
  year = 2025,
  doi = {10.48550/arXiv.2508.12939},
  archiveprefix = {arXiv}
}

@article{Papamakarios2016,
  title={Fast \$\epsilon\$-free Inference of Simulation Models with Bayesian Conditional Density Estimation},
  author={George Papamakarios and Iain Murray},
  journal={arXiv: Machine Learning},
  year={2016},
  url={https://api.semanticscholar.org/CorpusID:8401036}
}

@inproceedings{Greenberg2019,
  title={Automatic Posterior Transformation for Likelihood-Free Inference},
  author={David S. Greenberg and Marcel Nonnenmacher and Jakob H. Macke},
  booktitle={International Conference on Machine Learning},
  year={2019},
  url={https://api.semanticscholar.org/CorpusID:159041084}
}

@article{WC2020,
  title={Hierarchical Inference with Bayesian Neural Networks: An Application to Strong Gravitational Lensing},
  author={Sebastian Wagner-Carena and Ji Won Park and Simon Birrer and Philip J Marshall and Aaron Roodman and Risa Wechsler},
  journal={The Astrophysical Journal},
  year={2020},
  volume={909},
  url={https://api.semanticscholar.org/CorpusID:225075833}
}

@article{Uria2016NeuralAD,
  title={Neural Autoregressive Distribution Estimation},
  author={Benigno Uria and Marc-Alexandre C{\^o}t{\'e} and Karol Gregor and Iain Murray and H. Larochelle},
  journal={J. Mach. Learn. Res.},
  year={2016},
  volume={17},
  pages={205:1-205:37},
  url={https://api.semanticscholar.org/CorpusID:327844}
}

@article{JimenezRezende2015,
  title={Variational Inference with Normalizing Flows},
  author={Danilo Jimenez Rezende and Shakir Mohamed},
  journal={ArXiv},
  year={2015},
  volume={abs/1505.05770},
  url={https://api.semanticscholar.org/CorpusID:12554042}
}

@InProceedings{GG2015,
  title = 	 {MADE: Masked Autoencoder for Distribution Estimation},
  author = 	 {Germain, Mathieu and Gregor, Karol and Murray, Iain and Larochelle, Hugo},
  booktitle = 	 {Proceedings of the 32nd International Conference on Machine Learning},
  pages = 	 {881--889},
  year = 	 {2015},
  editor = 	 {Bach, Francis and Blei, David},
  volume = 	 {37},
  series = 	 {Proceedings of Machine Learning Research},
  address = 	 {Lille, France},
  month = 	 {07--09 Jul},
  publisher =    {PMLR},
  url = 	 {https://proceedings.mlr.press/v37/germain15.html}
}

@inproceedings{Hermans2019,
  title={Likelihood-free MCMC with Amortized Approximate Ratio Estimators},
  author={Joeri Hermans and Volodimir Begy and Gilles Louppe},
  booktitle={International Conference on Machine Learning},
  year={2019},
  url={https://api.semanticscholar.org/CorpusID:211132894}
}

@inproceedings{Papamakarios2018SequentialNL,
  title={Sequential Neural Likelihood: Fast Likelihood-free Inference with Autoregressive Flows},
  author={George Papamakarios and David C. Sterratt and Iain Murray},
  booktitle={International Conference on Artificial Intelligence and Statistics},
  year={2018},
  url={https://api.semanticscholar.org/CorpusID:29166658}
}

@ARTICLE{NPE-B,
       author = {{Lueckmann}, Jan-Matthis and {Bassetto}, Giacomo and {Karaletsos}, Theofanis and {Macke}, Jakob H.},
        title = "{Likelihood-free inference with emulator networks}",
      journal = {arXiv e-prints},
         year = 2018,
        month = may,
          eid = {arXiv:1805.09294},
        pages = {arXiv:1805.09294},
          doi = {10.48550/arXiv.1805.09294},
archivePrefix = {arXiv},
       eprint = {1805.09294},
 primaryClass = {stat.ML},
       adsurl = {https://ui.adsabs.harvard.edu/abs/2018arXiv180509294L}
}

@ARTICLE{Hermans2021ATC,
       author = {{Hermans}, Joeri and {Delaunoy}, Arnaud and {Rozet}, Fran{\c{c}}ois and {Wehenkel}, Antoine and {Begy}, Volodimir and {Louppe}, Gilles},
        title = "{A Trust Crisis In Simulation-Based Inference? Your Posterior Approximations Can Be Unfaithful}",
      journal = {arXiv e-prints},
         year = 2021,
        month = oct,
          eid = {arXiv:2110.06581},
        pages = {arXiv:2110.06581},
          doi = {10.48550/arXiv.2110.06581},
archivePrefix = {arXiv},
       eprint = {2110.06581},
 primaryClass = {stat.ML},
       adsurl = {https://ui.adsabs.harvard.edu/abs/2021arXiv211006581H}
}

@ARTICLE{DUNE1,
       author = {{Acciarri}, R. and others},
       collaboration = {DUNE collaboration},
        title = "{Long-Baseline Neutrino Facility (LBNF) and Deep Underground Neutrino Experiment (DUNE) Conceptual Design Report Volume 2: The Physics Program for DUNE at LBNF}",
      journal = {arXiv e-prints},
         year = 2015,
        month = dec,
          eid = {arXiv:1512.06148},
        pages = {arXiv:1512.06148},
          doi = {10.48550/arXiv.1512.06148},
archivePrefix = {arXiv},
       eprint = {1512.06148},
 primaryClass = {physics.ins-det},
       adsurl = {https://ui.adsabs.harvard.edu/abs/2015arXiv151206148D}
}

@ARTICLE{DUNE2,
       author = {{Abi}, B. and others},
       collaboration = {DUNE collaboration},
        title = "{Deep Underground Neutrino Experiment (DUNE), Far Detector Technical Design Report, Volume II: DUNE Physics}",
      journal = {arXiv e-prints},
         year = 2020,
        month = feb,
          eid = {arXiv:2002.03005},
        pages = {arXiv:2002.03005},
          doi = {10.48550/arXiv.2002.03005},
archivePrefix = {arXiv},
       eprint = {2002.03005},
 primaryClass = {hep-ex},
       adsurl = {https://ui.adsabs.harvard.edu/abs/2020arXiv200203005A}
}

@misc{lemos2023samplingbasedaccuracytestingposterior,
      title={Sampling-Based Accuracy Testing of Posterior Estimators for General Inference},
      author={Pablo Lemos and Adam Coogan and Yashar Hezaveh and Laurence Perreault-Levasseur},
      year={2023},
      eprint={2302.03026},
      archivePrefix={arXiv},
      primaryClass={stat.ML},
      url={https://arxiv.org/abs/2302.03026},
}

@article{ppp,
  author = "Abe, S. and others",
  title = "{Improving neutrino-nuclei interaction models: Recommendations and case studies on Peelle's Pertinent Puzzle}",
  eprint = "2509.17945",
  archivePrefix = "arXiv",
  primaryClass = "hep-ex",
  reportNumber = "FERMILAB-PUB-25-0691-CSAID",
  doi = "10.1103/615w-kjk2",
  journal = "Phys. Rev. D",
  volume = "113",
  number = "1",
  pages = "012011",
  year = "2026"
}

@article{Wolfs:2025ofb,
  author = "Wolfs, Jean and Marshall, Chris M.",
  title = "{Event generator tuning as a robustness test}",
  eprint = "2509.22526",
  archivePrefix = "arXiv",
  primaryClass = "hep-ph",
  doi = "10.1103/9dxd-871x",
  journal = "Phys. Rev. D",
  volume = "112",
  number = "9",
  pages = "092018",
  year = "2025"
}

@article{Hyper-Kamiokande:2018ofw,
    author = "Abe, K. and others",
    collaboration = "Hyper-Kamiokande",
    title = "{Hyper-Kamiokande Design Report}",
    eprint = "1805.04163",
    archivePrefix = "arXiv",
    primaryClass = "physics.ins-det",
    month = "5",
    year = "2018",
    journal = ""
}

@article{NuSTEC:2017hzk,
    author = "Alvarez-Ruso, L. and others",
    collaboration = "NuSTEC",
    title = "{NuSTEC White Paper: Status and challenges of neutrino{\textendash}nucleus scattering}",
    eprint = "1706.03621",
    archivePrefix = "arXiv",
    primaryClass = "hep-ph",
    reportNumber = "FERMILAB-PUB-17-195-ND-T, INT-PUB-17-020",
    doi = "10.1016/j.ppnp.2018.01.006",
    journal = "Prog. Part. Nucl. Phys.",
    volume = "100",
    pages = "1--68",
    year = "2018"
}

@article{Coyle:2025xjk,
    author = "Coyle, Nina M. and Li, Shirley Weishi and Machado, Pedro A. N.",
    title = "{Neutrino-nucleus cross section impacts on neutrino oscillation measurements}",
    eprint = "2502.19467",
    archivePrefix = "arXiv",
    primaryClass = "hep-ph",
    reportNumber = "FERMILAB-PUB-25-0001-T, UCI-HEP-TR-2025-01",
    doi = "10.1103/PhysRevD.111.093010",
    journal = "Phys. Rev. D",
    volume = "111",
    number = "9",
    pages = "093010",
    year = "2025"
}

@inproceedings{Tame-Narvaez:2025pwg,
    author = "Tame-Narvaez, Karla and {\'C}iprijanovi{\'c}, Aleksandra and Gardiner, Steven and Cerati, Giuseppe",
    title = "{Simulation-Based Inference for Neutrino Interaction Model Parameter Tuning}",
    booktitle = "{39th Annual Conference on Neural Information Processing Systems}: {Includes Machine Learning and the Physical Sciences (ML4PS)}",
    eprint = "2510.07454",
    archivePrefix = "arXiv",
    primaryClass = "hep-ph",
    reportNumber = "FERMILAB-CONF-25-0614-CSAID-PPD-T",
    month = "10",
    year = "2025"
}

@article{MicroBooNE:2025nll,
    author = "Abratenko, P. and others",
    collaboration = "MicroBooNE",
    title = "{Search for light sterile neutrinos with two neutrino beams at MicroBooNE}",
    eprint = "2512.07159",
    archivePrefix = "arXiv",
    primaryClass = "hep-ex",
    reportNumber = "FERMILAB-PUB-24-0865-PPD",
    doi = "10.1038/s41586-025-09757-7",
    journal = "Nature",
    volume = "648",
    number = "8092",
    pages = "64--69",
    year = "2025"
}

@article{MicroBooNE:2025aiw,
    author = "Abratenko, P. and others",
    collaboration = "MicroBooNE",
    title = "{Measurements of differential charged-current cross sections on argon for electron neutrinos with final-state protons in MicroBooNE}",
    eprint = "2511.17342",
    archivePrefix = "arXiv",
    primaryClass = "hep-ex",
    reportNumber = "FERMILAB-PUB-25-0625-PPD",
    month = "11",
    year = "2025",
    journal = ""
}

@Article{golan2012,
  author	= {T. Golan and J.T. Sobczyk and J. Żmuda},
  title		= {{NuWro}: the {W}rocław {Monte Carlo} Generator of
		  Neutrino Interactions},
  journal	= {Nucl. Phys. B - Proc. Suppl.},
  volume	= {229-232},
  pages		= {499},
  year		= {2012},
  issn		= {0920-5632},
  doi		= {10.1016/j.nuclphysbps.2012.09.136}
}

@inproceedings{fruhwirth2012peelle,
  title={Peelle’s pertinent puzzle and its solution},
  author={Fr{\"u}hwirth, R and Neudecker, D and Leeb, H},
  booktitle={EPJ Web of Conferences},
  volume={27},
  pages={00008},
  year={2012},
  organization={EDP Sciences},
  doi={10.1051/epjconf/20122700008}
}

@article{Koch:2021yda,
    author = "Koch, Lukas",
    title = "{Robust test statistics for data sets with missing correlation information}",
    eprint = "2102.06172",
    archivePrefix = "arXiv",
    primaryClass = "physics.data-an",
    doi = "10.1103/PhysRevD.103.113008",
    journal = "Phys. Rev. D",
    volume = "103",
    number = "11",
    pages = "113008",
    year = "2021"
}

@article{Chakrani:2023htw,
    author = "Chakrani, J. and others",
    title = "{Parametrized uncertainties in the spectral function model of neutrino charged-current quasielastic interactions for oscillation analyses}",
    eprint = "2308.01838",
    archivePrefix = "arXiv",
    primaryClass = "hep-ex",
    doi = "10.1103/PhysRevD.109.072006",
    journal = "Phys. Rev. D",
    volume = "109",
    number = "7",
    pages = "072006",
    year = "2024"
}

@article{Juszczak:2009qa,
    author = "Juszczak, Cezary",
    editor = "Ankowski, Arthur and Sobczyk, Jan",
    title = "{Running NuWro}",
    eprint = "0909.1492",
    archivePrefix = "arXiv",
    primaryClass = "hep-ex",
    journal = "Acta Phys. Polon. B",
    volume = "40",
    pages = "2507--2512",
    year = "2009"
}
%%%%%%%%%%%%%%%%%%%%%%%%%%%%%%%%%%%%

% \clearpage
\appendix
\section{Additional bins from the T2K data set}\label{appA}
Figure~\ref{fig:nuwro_all} presents the remaining bins of the two-dimensional ``Analysis~I'' T2K cross-section measurement~\cite{T2K:2016jor} that are included in our analysis but not shown in Figure~\ref{fig:nuwro_main}. Each panel shows a single bin of scattering cosine $\cos\theta$, which is subdivided into bins of muon momentum $p_\mu$. The plotting format is identical to Figure~\ref{fig:nuwro_main}. In these portions of the data set, the inferred predictions follow trends similar to those already discussed in the main text.
\begin{figure*}[htb!]
\centering

\begin{subfigure}{0.45\textwidth}
    \centering
    \includegraphics[width=\linewidth]{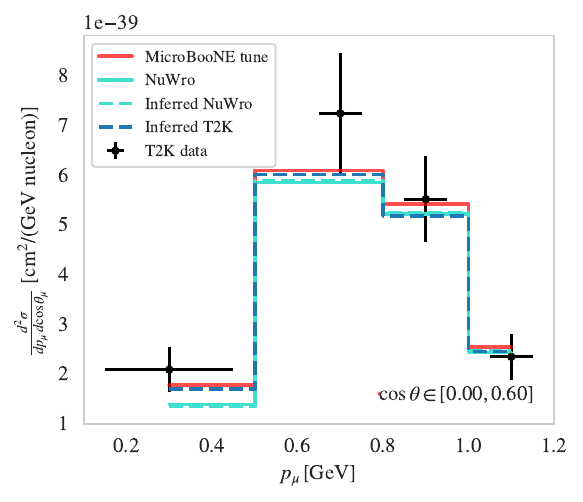}
    %\caption{NuWro slice 1}
    \label{fig:nuwro2}
\end{subfigure}
\begin{subfigure}{0.45\textwidth}
    \centering
    \includegraphics[width=\linewidth]{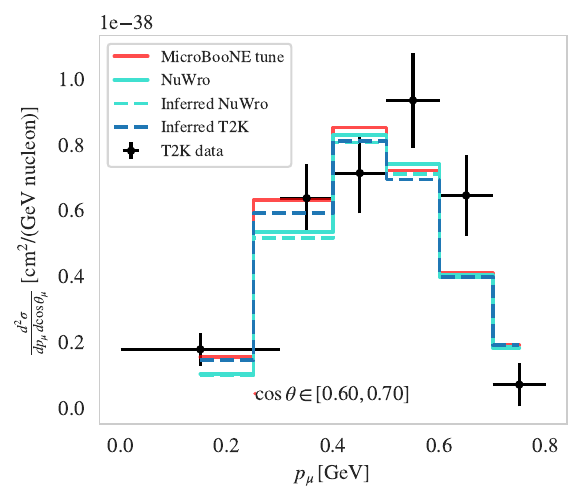}
    %\caption{NuWro slice 2}
    \label{fig:nuwro2}
\end{subfigure}
% \hfill
\begin{subfigure}{0.45\textwidth}
    \centering
    \includegraphics[width=\linewidth]{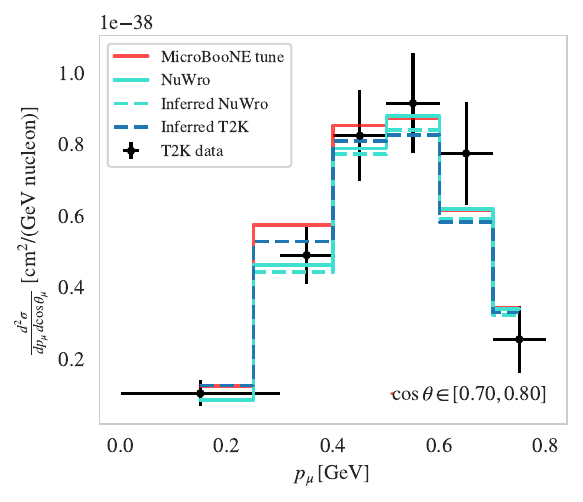}
    %\caption{NuWro slice 3}
    \label{fig:nuwro3}
\end{subfigure}
\begin{subfigure}{0.45\textwidth}
    \centering
    \includegraphics[width=\linewidth]{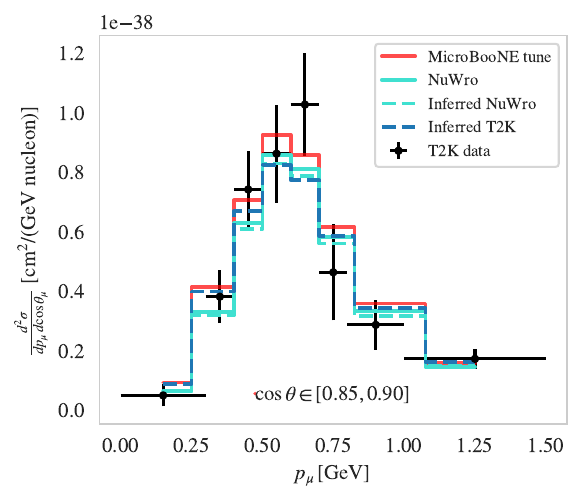}
    %\caption{NuWro slice 5}
    \label{fig:nuwro5}
\end{subfigure}
\begin{subfigure}{0.45\textwidth}
    \centering
    \includegraphics[width=\linewidth]{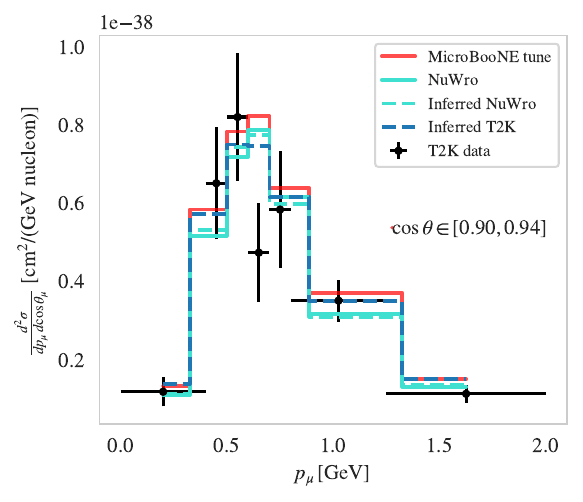}
    %\caption{NuWro slice 6}
    \label{fig:nuwro6}
\end{subfigure}
\begin{subfigure}{0.45\textwidth}
    \centering
    \includegraphics[width=\linewidth]{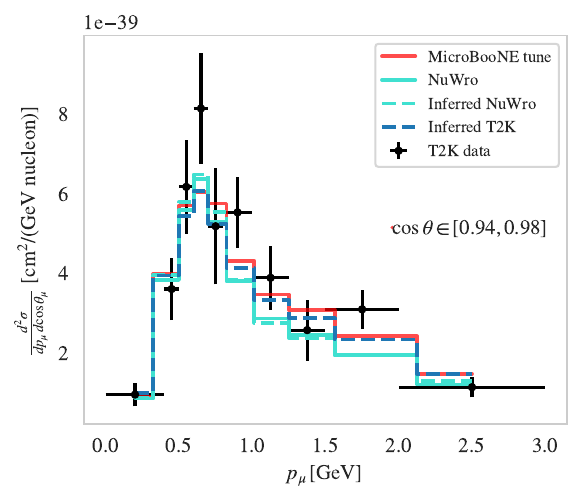}
    %\caption{NuWro slice 7}
    \label{fig:nuwro7}
\end{subfigure}

\caption{Comparison of the T2K cross-section data (black points) to the NuWro~\cite{Juszczak:2009qa} (solid cyan), NPE-inferred NuWro (dashed cyan), MicroBooNE Tune~\cite{microboonegenietune} (solid red), and NPE-inferred T2K (dashed blue) model predictions in different slices of $\cos\theta$.}
\label{fig:nuwro_all}
\end{figure*}

\end{document}